\documentclass[reviewcopy]{elsart}
\usepackage{graphicx}
\usepackage{amssymb}


\begin{document}

\begin{frontmatter}

\title{Universality of the Collins-Soper-Sterman nonperturbative 
function\\ in vector boson production}

\author[Anton]{Anton V. Konychev}
 and \author[Pavel]{Pavel M. Nadolsky}

\address[Anton]{Department of Physics, Indiana University, Bloomington, IN 47405-7105, U.S.A.}
\address[Pavel]{High Energy Physics Division, Argonne National Laboratory, 
\\ Argonne, IL 60439-4815, U.S.A.}

\date{\today{}}

\begin{abstract}
We revise the $b_*$ model for the
Collins-Soper-Sterman resummed form factor to improve description of
the leading-power contribution at nearly nonperturbative impact parameters. 
This revision leads to excellent agreement of the transverse momentum
resummation with the data in a global analysis of
Drell-Yan lepton pair and $Z$ boson production. The nonperturbative 
contributions are found to follow universal quasi-linear dependence 
on the logarithm of the heavy boson invariant mass, which closely 
agrees with an estimate from the infrared renormalon analysis.
\end{abstract}
\end{frontmatter}
Transverse momentum distributions of heavy Drell-Yan lepton pairs,
$W$, or $Z$ bosons produced in hadron-hadron collisions present an
interesting example of factorization for multi-scale observables.
If the transverse momentum $q_{T}$ of the electroweak boson is much
smaller than its invariant mass $Q$, $d\sigma/dq_{T}$ at an n-th order of perturbation theory 
includes large contributions of the type $\alpha_{s}^{n}{\mathrm{ln}}^{m}(q_{T}^{2}/Q^{2})/q_{T}^{2}$
($m=0,1\ldots2n-1$), which must be summed through all orders of $\alpha_{s}$
to reliably predict the cross section \cite{PP}.
The feasibility of all-order resummation is proved by a factorization
theorem, first formulated for $e^{+}e^{-}$ hadroproduction \cite{CS1,CS2},
stated by Collins, Soper, and Sterman (CSS) for the Drell-Yan process
\cite{CSS}, and recently proved by detailed investigation of gauge
transformations of $k_{T}$-dependent parton densities \cite{CM,JiMaYuan}. 

The heavy bosons acquire non-zero $q_T$ mostly by recoiling 
against QCD radiation. The CSS formalism accounts for both 
the short- and long-wavelength QCD radiation 
by means of a Fourier-Bessel transform of a resummed
form factor $\widetilde{W}(b)$ introduced in impact parameter ($b$) space. 
The perturbative contribution, characterized by 
$b \lesssim 0.5\mbox{ GeV}^{-1}$, dominates in $W$ and $Z$ boson
production at all values of $q_{T}$. 
The nonperturbative contribution from $b\gtrsim 0.5\mbox{ GeV}^{-1}$
is not negligible at $q_{T}<20$ GeV in the precision measurements of
the $W$ boson mass $M_{W}$ at the Tevatron and LHC~\cite{N}.
The model for the nonperturbative recoil is the major source of theoretical
uncertainty in the extraction of $M_{W}$ from the experimental data.
This uncertainty must be reduced in order to measure $M_{W}$ with 
accuracy of about 30 MeV in the Tevatron Run-2 and 15 MeV at the LHC. 
The nonperturbative model presented below approaches the 
level of accuracy desired in these measurements.

The nonperturbative component [described by the function
${\mathcal F}_{NP}(b,Q)$ given in Eq.\,(\ref{W2})] can be constrained
in a few experiments by exploiting process-independence, or universality, of ${\mathcal F}_{NP}(b,Q)$,
just as the universal $k_{T}$-integrated parton densities are constrained
with the help of inclusive scattering data. The universality 
of ${\mathcal F}_{NP}(b,Q)$
in unpolarized Drell-Yan-like processes and semi-inclusive deep-inelastic
scattering (SIDIS) follows from the CSS factorization theorem \cite{CM}. 
In the study presented here, 
we carefully investigate agreement of the universality 
assumption with the data in a global analysis of fixed-target 
Drell-Yan pair and Tevatron $Z$ boson production.
We revise the nonperturbative model used in the previous studies 
\cite{DWSLYERV,BLNY} and improve agreement with the data without
introducing additional free parameters. Renormalization-group 
invariance requires ${\mathcal F}_{NP}(b,Q)$ to depend linearly 
on $\ln Q$ \cite{CS2,CSS}. With our latest revisions put in place, 
the global $q_T$ fit clearly prefers a simple
function ${\mathcal F}_{NP}(b,Q)$ with universal $\ln Q$
dependence. The new ${\mathcal F}_{NP}(b,Q)$ has reduced dependence on the 
collision energy $\sqrt{S}$ comparatively to the earlier fits. 
The slope of the $\ln Q$ dependence found 
in the new fit agrees numerically with its estimate 
made with methods of infrared renormalon analysis \cite{KorS,Tafat}.

The function ${\mathcal F}_{NP}(b,Q)$ primarily
parametrizes the ``power-suppressed'' terms, {\it i.e.}, terms 
proportional to positive powers of $b$. When assessed in a 
fit, ${\mathcal F}_{NP}(b,Q)$ also contains admixture of
the leading-power terms (logarithmic in $b$ terms), which were not 
properly included in the approximate leading-power function
$\widetilde{W}_{LP}(b)$ [cf. Eq.~(\ref{W2})]. 
In contrast, estimates of  ${\mathcal F}_{NP}(b,Q)$ 
made in the infrared renormalon analysis explicitly remove 
all leading-power contributions from 
${\mathcal  F}_{NP}(b,Q)$ \cite{Tafat}.
While the recent studies \cite{BLNY,KorS,Tafat,QZ,KSV} point 
to an approximately Gaussian form of ${\mathcal F}_{NP}(b,Q)$ {[}${\mathcal
    F}_{NP}(b,Q) \propto b^2${]}, they disagree on the  
magnitude of ${\mathcal F}_{NP}(b,Q)$ and its $Q$ dependence. 
The source of these differences 
can be traced to the varying assumptions about the form
of the leading-power function $\widetilde{W}_{LP}(b)$  
at $b<2\mbox{ GeV}^{-1}$, 
which is correlated in the fit with ${\mathcal F}_{NP}(b,Q)$. 
The exact behavior of $\widetilde{W}(b)$ at 
$b>2\mbox{ GeV}^{-1}$ is of reduced importance,
as $\widetilde{W}(b)$ is strongly suppressed at such $b$.
The new improvements described here (excellent agreement of ${\mathcal
  F}_{NP}(b,Q)$ with the data and renormalon analysis)
result from modifications in the model for 
$\widetilde{W}_{LP}(b)$ at $b<2\mbox{ GeV}^{-1}$. 
The improvements are preserved under variations of the large-$b$ form 
of $\widetilde{W}_{LP}(b)$ in a significant range of the model parameters.


Our paper follows the notations in Ref.~\cite{BLNY}. The form factor 
$\widetilde{W}(b)$ factorizes at all $b$ as \cite{CS1,CS2,CSS}
\begin{equation}
\widetilde{W}(b)=\sum_{j=q,\bar{q}}\frac{\sigma_{j}^{(0)}}{S}\, 
e^{-{\mathcal{S}}(b,Q)}{\mathcal{P}}_{j}(x_{1},b)
{\mathcal{P}}_{\bar{j}}(x_{2},b),\label{W}
\end{equation}
where $\sigma_{j}^{(0)}/S$ is a constant prefactor \cite{CSS}, 
and $x_{1,2}\equiv e^{\pm  y}Q/\sqrt{S}$ are the Born-level momentum 
fractions, with $y$ being the rapidity of the vector boson. 
The $b$-dependent parton densities ${\mathcal{P}}_{j}(x,b)$ and 
Sudakov function \begin{eqnarray}
{\mathcal{S}}(b,Q)\equiv\int_{b_{0}^{2}/b^{2}}^{Q^{2}}\frac{d\bar{\mu}^{2}}
{\bar{\mu}^{2}}\biggl[{\mathcal{A}}(\alpha_{s}(\bar{\mu}))\,
\mathrm{ln}\biggl(\frac{Q^{2}}{\bar{\mu}^{2}}\biggr)+{\mathcal{B}}(\alpha_{s}(\bar{\mu}))\biggr]\end{eqnarray}
are universal in Drell-Yan-like processes and SIDIS \cite{CM}. When the 
momentum scales $Q$ and $b_{0}/b$ (where $b_{0}\equiv
2e^{-\gamma_{E}}\approx1.123$ is a dimensionless constant)
are much larger than 1 GeV,
$\widetilde{W}(b)$ reduces to its perturbative part $\widetilde{W}_{pert}(b)$,
{\it i.e.}, its leading-power (logarithmic in $b$) part evaluated at a finite order 
of $\alpha_s$:
\begin{eqnarray}
\left.\widetilde{W}(b)\right|_{Q,b_{0}/b\gg1\, GeV}\approx\widetilde{W}_{pert}(b)\equiv\sum_{j=q,\bar{q}}\frac{\sigma_{j}^{(0)}}{S}\, e^{-{\mathcal{S}}_{P}(b,Q)}\nonumber \\
\times\,\left[{\mathcal C}\otimes f\right]_{j}(x_{1},b;\,
\mu_{F})\left[{\mathcal C}\otimes f\right]_{\bar{j}}(x_{2},b;\,\mu_{F}).\label{Wpert}\end{eqnarray}
Here ${\mathcal S}_{P}(b,Q)$ and $
\left[{\mathcal C}\otimes f\right]_{j}(x,b;\,\mu_{F})\equiv\sum_{a}{\int_{x}^{1}{d\xi/\xi\,{\mathcal{C}}_{ja}(x/\xi,\mu_{F}b)}}f_{a}(\xi,\mu_{F})$
are the finite-order approximations to the leading-power parts of ${\mathcal{S}}(b,Q)$ and ${\mathcal{P}}_{j}(x,b)$. $f_{a}(x,\mu_{F})$ is the $k_{T}$-integrated
parton density, computed in our study by using the CTEQ6M parameterization
\cite{CTEQ6}. ${\mathcal{C}}_{ja}(x,\mu_{F}b)$ is the Wilson
coefficient function. We compute  ${\mathcal S}_{P}(b,Q)$ up to
$O(\alpha_{s}^{2})$
and ${\mathcal{C}}_{ja}$ up to $O(\alpha_{s})$. 

In $Z$ boson production, the maximum of $b\widetilde{W}(b)$
is located at $b\approx0.25\mbox{ GeV}^{-1}$, and $\widetilde{W}_{pert}(b)$
dominates the Fourier-Bessel integral.
In the examined low-$Q$ region, the maximum of $b\widetilde{W}(b)$
is located at $b\approx1\mbox{ GeV}^{-1}$, where higher-order corrections
in powers of $\alpha_{s}$ and $b$ must be considered. 
We reorganize  Eq.~(\ref{W}) to 
separate the leading-power (LP) term $\widetilde{W}_{LP}(b)$, given by
the model-dependent continuation of $\widetilde{W}_{pert}(b)$ to
$b\gtrsim1\mbox{ GeV}^{-1}$, and the nonperturbative exponent $e^{-{\mathcal F}_{NP}(b,Q)}$,
which absorbs the power-suppressed terms: 
\begin{equation}
\widetilde{W}(b)=\widetilde{W}_{LP}(b)\, 
e^{-{\mathcal {\mathcal F}}_{NP}(b,Q)}.
\label{W2}
\end{equation}
At $b\rightarrow0$, the perturbative approximation for $\widetilde{W}(b)$
is restored: $\widetilde{W}_{LP}\rightarrow\widetilde{W}_{pert},$
${\mathcal F}_{NP}\rightarrow0$. The power-suppressed contributions are
proportional to even powers of $b$ \cite{KorS}. Detailed
expressions for some power-suppressed terms are given in Ref.~\cite{Tafat}. At 
impact parameters of order 1$\mbox{ GeV}^{-1}$, we
keep only the first power-suppressed contribution proportional to
$b^2$:\begin{equation}
{\mathcal F}_{NP}\approx
b^{2}\left(a_{1}+a_{2}\ln(Q/Q_{0})+a_{3}\phi(x_{1})+
a_{3}\phi(x_{2})\right)+...\,,\label{FNP1}\end{equation}
where $a_1$, $a_2$, and $a_3$ are coefficients 
of magnitude less than $1\mbox{ GeV}^2$, and $\phi(x)$ is a
dimensionless function.
The terms $a_{2}\ln(Q/Q_{0})$ and $a_{3}\phi(x_{j})$ arise from ${\mathcal S}(b,Q)$ and
$\ln\bigl[{\mathcal P}_{j}(x_{j},b)\bigr]$ in $\ln\bigl[\widetilde{W}(b)\bigr]$, respectively. 
We neglect the flavor dependence of  $\phi(x)$ 
in the analyzed region dominated by
scattering of light $u$ and $d$ quarks. ${\mathcal F}_{NP}$ is
consequently a universal function within this region. The dependence
of ${\mathcal F}_{NP}$ on $\ln Q$ follows from renormalization-group
invariance of the soft-gluon radiation \cite{CS2}. The coefficient $a_{2}$
of the $\ln Q$ term has been related to the vacuum average of the
Wilson loop operator and estimated within lattice QCD as
$0.19_{-0.09}^{+0.12}\mbox{ GeV}^{2}$\nolinebreak[4] 
\cite{Tafat}. 

The preferred ${\mathcal F}_{NP}$ is correlated in the fit with
the assumed large-$b$ behavior of $\widetilde{W}_{LP}$. We examine this correlation in a modified
version of the $b_{*}$ model \cite{CS2,CSS}. The shape of $\widetilde{W}_{LP}$
is varied in the $b_*$ model by adjusting a single parameter
$b_{max}$. Continuity of $\widetilde{W}$ and its derivatives,
needed for the numerical stability of the Fourier transform, is always preserved.
We set $\widetilde{W}_{LP}(b)\equiv\widetilde{W}_{pert}(b_{*}),$
with $b_{*}(b,b_{max})\equiv b(1+b^{2}/b_{max}^{2})^{-1/2}$. $\widetilde{W}_{LP}(b)$
reduces to $\widetilde{W}_{pert}(b)$ as $b\rightarrow0$ and asymptotically
approaches $\widetilde{W}_{pert}(b_{max})$ as $b\rightarrow\infty$.
The $b_{*}$ model with a relatively low $b_{max}=0.5\mbox{ GeV}^{-1}$
was a choice of the previous $q_{T}$ fits \cite{DWSLYERV,BLNY}.
However, it is natural to consider $b_{max}$ above 1$\mbox{ GeV}^{-1}$
in order to avoid \emph{ad hoc} modifications of $\widetilde{W}_{pert}(b)$
in the region where perturbation theory is still applicable. 
To implement
$\widetilde{W}_{pert}(b_*)$ for $b_{max}>1\mbox{ GeV}^{-1}$, we must 
choose the factorization scale $\mu_{F}$ such that it stays, at any $b$ 
and $b_{max}$, above the initial scale $Q_{ini}=1.3$ GeV of the DGLAP 
evolution for the CTEQ6 PDF's $f_{a}(x,\mu_{F})$. 
We keep the usual choice
$\mu_{F}=C_{3}/b_{*}(b,b_{max})$, where $C_3\sim b_0$, for $b_{max}\leq
b_{0}/Q_{ini} \approx 0.86 \mbox{ GeV}^{-1}$.
Such choice is not acceptable at $b_{max} > b_{0}/Q_{ini}$, as it would allow $\mu_F<Q_{ini}$. Instead,
we choose $\mu_{F}=C_{3}/b_{*}(b,b_{0}/Q_{ini})$ for
$b_{max}>b_{0}/Q_{ini}$, {\it i.e.}, we substitute $b_0/Q_{ini}$ 
for $b_{max}$ in $\mu_F$ to satisfy $\mu_F \geq Q_{ini}$ at any $b$.
Aside from $f_{a}(x,\mu_F)$, all terms in
$\widetilde{W}_{pert}(b)$ are known, at least formally, 
as explicit functions of $\alpha_s(1/b)$ at all $b < 1/\Lambda_{QCD}$. 
We show in
Ref.~\cite{KNlong} that this prescription preserves  correct resummation
of the large logarithms and is numerically stable up to $b_{max} \sim 3\mbox{ GeV}^{-1}$.

We perform a series of fits for several choices of $b_{max}$ by
closely following the previous global $q_{T}$
analysis \cite{BLNY}. We consider a total of 98 data points from
production of Drell-Yan pairs in E288, E605, and R209 fixed-target
experiments, as well as from observation of $Z$ bosons with $q_{T}<10$
GeV by CDF and D\O\, detectors in the Tevatron Run-1. See Ref.~\cite{BLNY} for
the experimental references. 
Overall normalizations
for the experimental cross sections are varied as free parameters.
Our best-fit normalizations agree with the published values within
the systematical errors provided by the experiments, with the exception
of the CDF Run-1 normalization (rescaled down by 7\%).

%
\begin{figure}[tb]
\begin{center}
\includegraphics[%
  width=10cm,
  keepaspectratio]{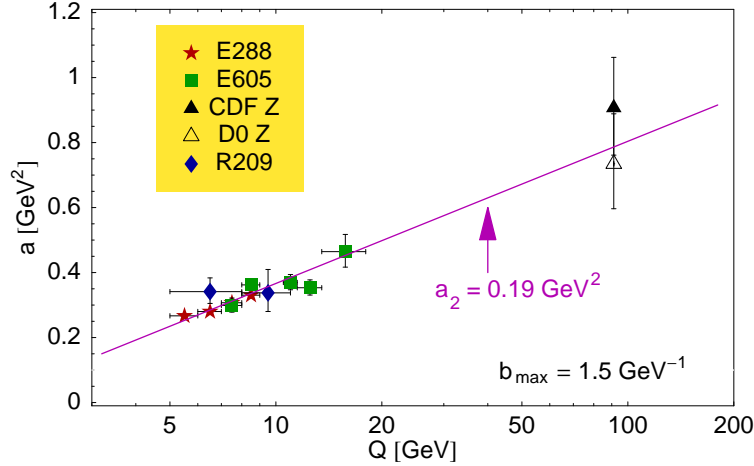}
\end{center}
\caption{The best-fit values of $a(Q)$ obtained
in independent scans of $\chi^{2}$ for the contributing experiments.
The vertical error bars correspond to the increase of $\chi^{2}$ by unity
above its minimum in each $Q$ bin. The slope of the line is equal
to the central-value prediction from the renormalon analysis \cite{Tafat}. 
\label{fig:aQ_vs_Q}}
\end{figure}

To test the universality of ${\mathcal F}_{NP}$, we individually
examine each bin of $Q$. We choose ${\mathcal F}_{NP} = a(Q)b^{2}$
and independently fit it to each of the 5 experimental data sets to determine
the most plausible normalization in each experiment. We then set the
normalizations equal to their best-fit values and examine $\chi^{2}$
at each $Q$ as a function of $a(Q)$. For $b_{max}=1-2\mbox{ GeV}^{-1}$,
the best-fit values of $a(Q)$ follow a nearly linear dependence on $\ln Q$ {[}cf. Fig.~\ref{fig:aQ_vs_Q}{]}.
The slope $a_{2}\equiv da(Q)/d(\ln Q)$ 
is close to the renormalon analysis expectation of $~0.19\mbox{ GeV}^{2}$
\cite{Tafat}. The agreement with the universal linear $\ln Q$ dependence
worsens if $b_{max}$ is chosen outside the region 1-2$\mbox{ GeV}^{-1}$.
Since the best-fit $a(Q)$ are found independently in each $Q$ bin,
we conclude that the data support the universality of ${\mathcal
  F}_{NP}$, when  $b_{max}$ lies in the range $1-2\mbox{ GeV}^{-1}$.
In another test, we find that each experimental data set
individually prefers a nearly quadratic dependence on $b$, 
${\mathcal F}_{NP} = a(Q) b^{2-\beta}$, with $|\beta| < 0.5$ 
in all five experiments.

To further explore the issue, we simultaneously fit our model
to all the data. We parametrize $a(Q)$ as $
a(Q)\equiv a_{1}+a_{2}\mathrm{ln}\left[Q/(3.2\mbox{ GeV})\right]+a_{3}\mathrm{ln}\left[100x_{1}x_{2}\right].$
This parametrization coincides with the BLNY form \cite{BLNY}, if
the parameters are renamed as 
$\{ g_{1},g_{2},g_{1}g_{3}\}\mbox{(BLNY)}\rightarrow\{
a_{1},a_{2},a_{3}\}\mbox{{(here)}}$. It agrees with the generic form of 
${\mathcal F}_{NP}(b,Q)$ in Eq.~(\ref{FNP1}), if one identifies ${\phi}(x)=\ln(x/0.1)$.
We carry out two sequences of fits 
for $C_3=b_0$ and $C_3=2b_0$ to investigate the stability 
of our prescription for $\mu_F$ and sensitivity to ${\mathcal O}(\alpha_s^2)$ 
corrections. The dependence on $C_3$ is relatively uniform across the
whole range of $b_{max}$, indicating that our choice of $\mu_F$ 
for $b_{max}>b_0/Q_{ini}$ is numerically stable. 

\begin{figure}[tb]
\begin{center}
\includegraphics[%
  width=11cm,
  keepaspectratio]{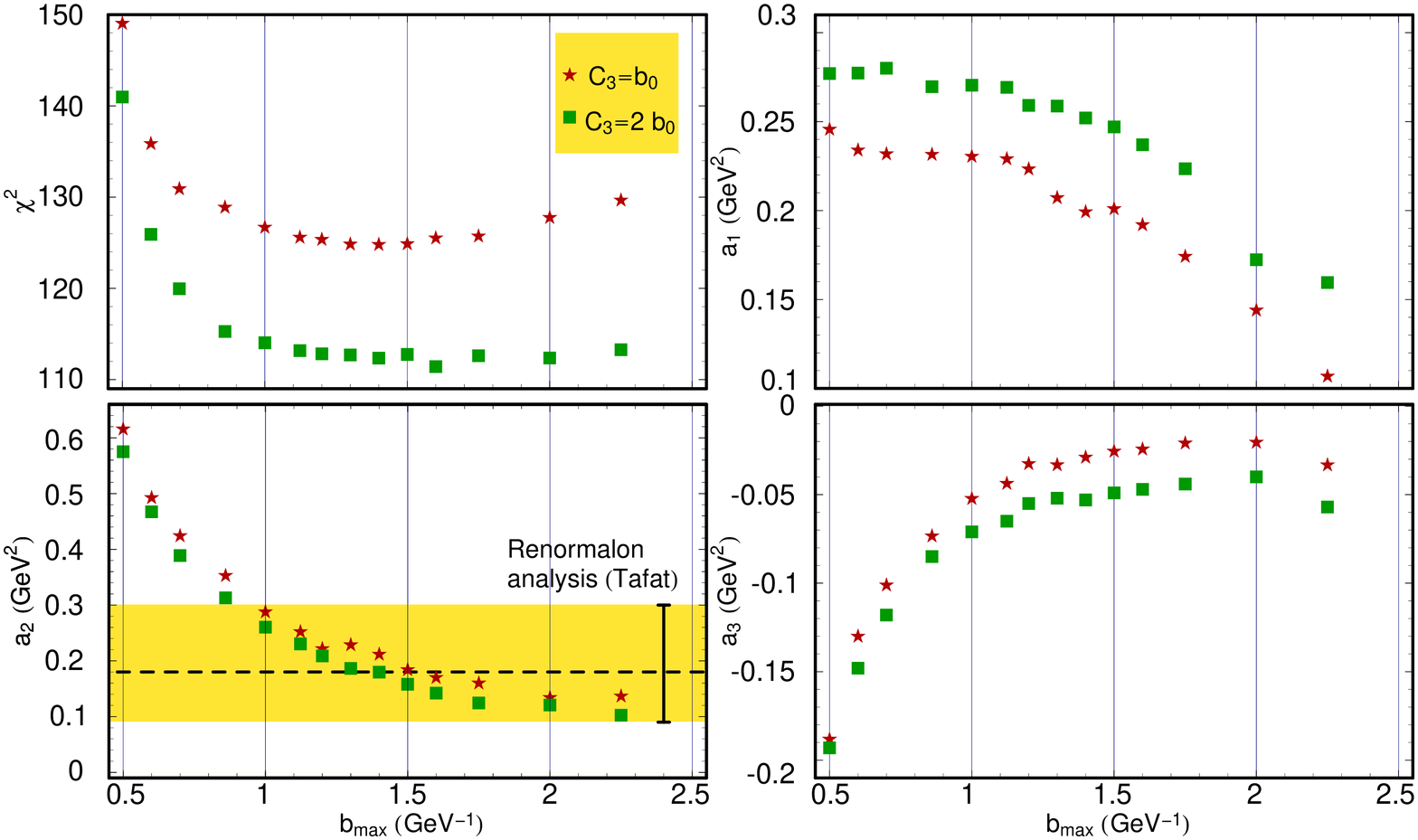}
\end{center}
\caption{The best-fit $\chi^{2}$ and coefficients $a_{1}$, $a_{2}$, and
$a_{3}$ in ${\mathcal F}_{NP}(b,Q)$ for different values of \protect $b_{max}$,
$C_{3}=b_{0}$ (stars) and $C_{3}=2b_{0}$ (squares). The size of
the symbols approximately corresponds to $1\sigma$ errors for the
shown parameters. \label{fig:chi2a123}}
\end{figure}

\begin{figure}[tb]
\begin{center}
\includegraphics[%
  height=6cm,
  keepaspectratio]{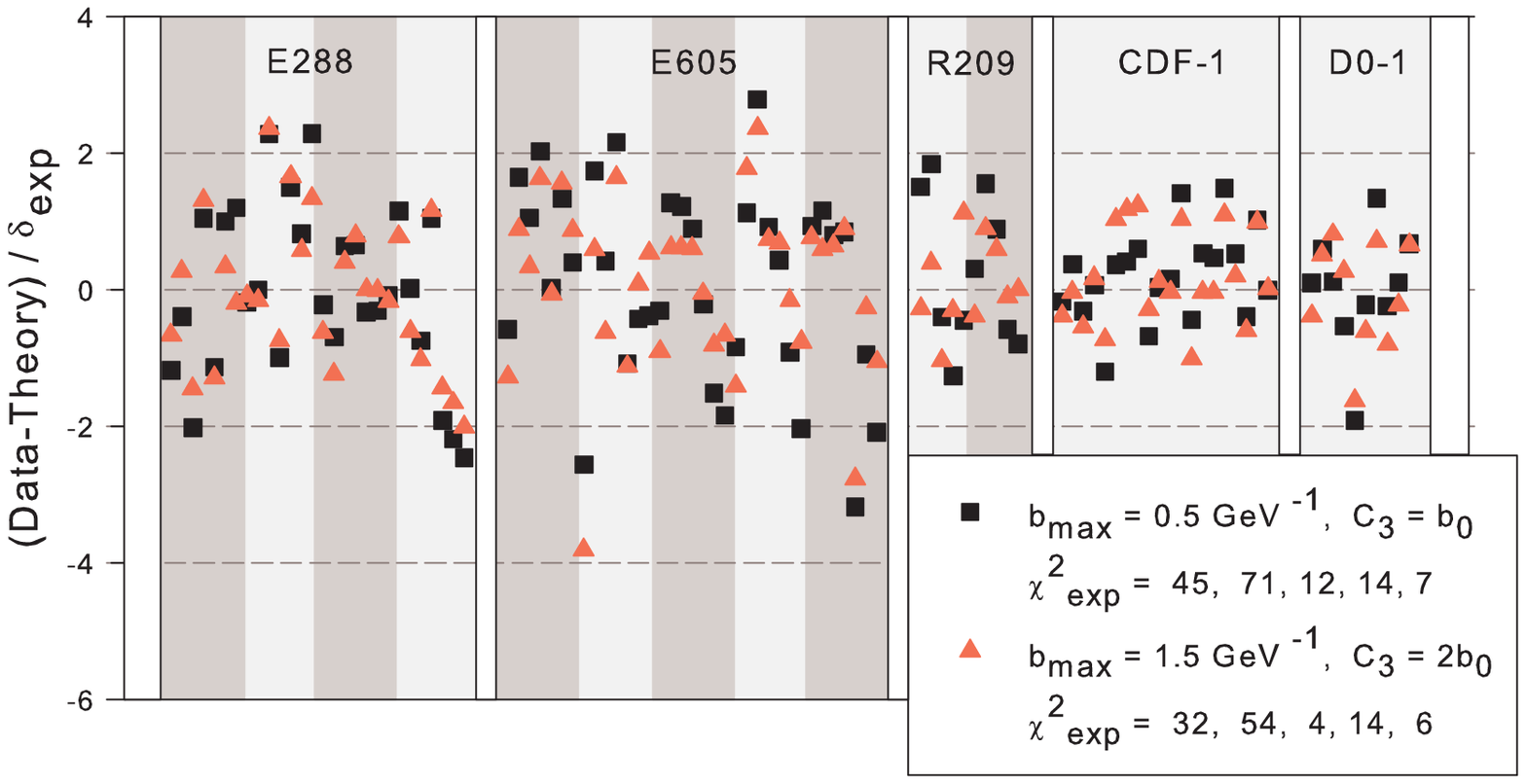}
\end{center}
\caption{Differences between the measured (Data) and theoretical (Theory) cross sections, divided by the experimental 
error $\delta_{exp}$ in each $(Q,q_T)$ bin. The values of $\chi^2$ for each experiment in the two fits are listed in the legend in the same order. \label{fig:fit_quality}}
\end{figure}

Fig.~\ref{fig:chi2a123} shows the dependence of the best-fit $\chi^{2},$
$a_{1},$ $a_{2}$, and $a_{3}$ on $b_{max}$. As $b_{max}$ is increased
above $0.5\mbox{ GeV}^{-1}$ assumed in the BLNY study, $\chi^{2}$
rapidly decreases, becomes relatively flat at $b_{max}=1-2\mbox{
  GeV}^{-1}$, and grows again at $b_{max} > 2 \mbox{
  GeV}^{-1}$. The global minimum of $\chi^{2}=125\,(111)$ 
is reached at $b_{max}\approx1.5$ GeV$^{-1}$, where all data sets 
are described equally well, without major tensions among the five
experiments. 
The improvement in $\chi^{2}$ mainly ensues from 
better agreement with the low-$Q$ experiments (E288, E605, and R209), 
while the quality of all fits to the $Z$ data is about the same.
This is illustrated by Fig.~\ref{fig:fit_quality}, which 
shows the differences between the measured 
and theoretical cross sections, divided by the experimental errors 
$\delta_{exp}$, as well as the values of $\chi^2$ in each experiment, 
in two representative fits for
$b_{max} = 0.5 \mbox{ GeV}^{-1}$, $C_{3}=b_{0}$ (squares) and 
$b_{max} = 1.5 \mbox{ GeV}^{-1}$, $C_{3}=2b_{0}$ (triangles).
The data are arranged in bins of $Q$ (shown by gray background stripes) 
and $q_T$, with both variables increasing from left to right. 
For $b_{max} =1.5 \mbox{ GeV}^{-1}$, the $(\mbox{Data}-\mbox{Theory})$ 
differences are  reduced on average in the entire low-$Q$ sample, 
resulting in lower $\chi^2$ in three low-$Q$ experiments. Two outlier points 
in the E605 sample (the first point
in the second $Q$ bin and fifth point in the fifth $Q$ bin) 
disagree with the other E288 and E605 data in the same $Q$ and $x$ region and 
contribute $15-25$ units to $\chi^{2}$ at any $b_{max}$. 
If the two outliers were removed, one would find $\chi^2/d.o.f.\approx
1$ at the global minimum.

\begin{figure}[tb]
\begin{center}
\includegraphics[%
  width=0.47\textwidth,
  keepaspectratio]{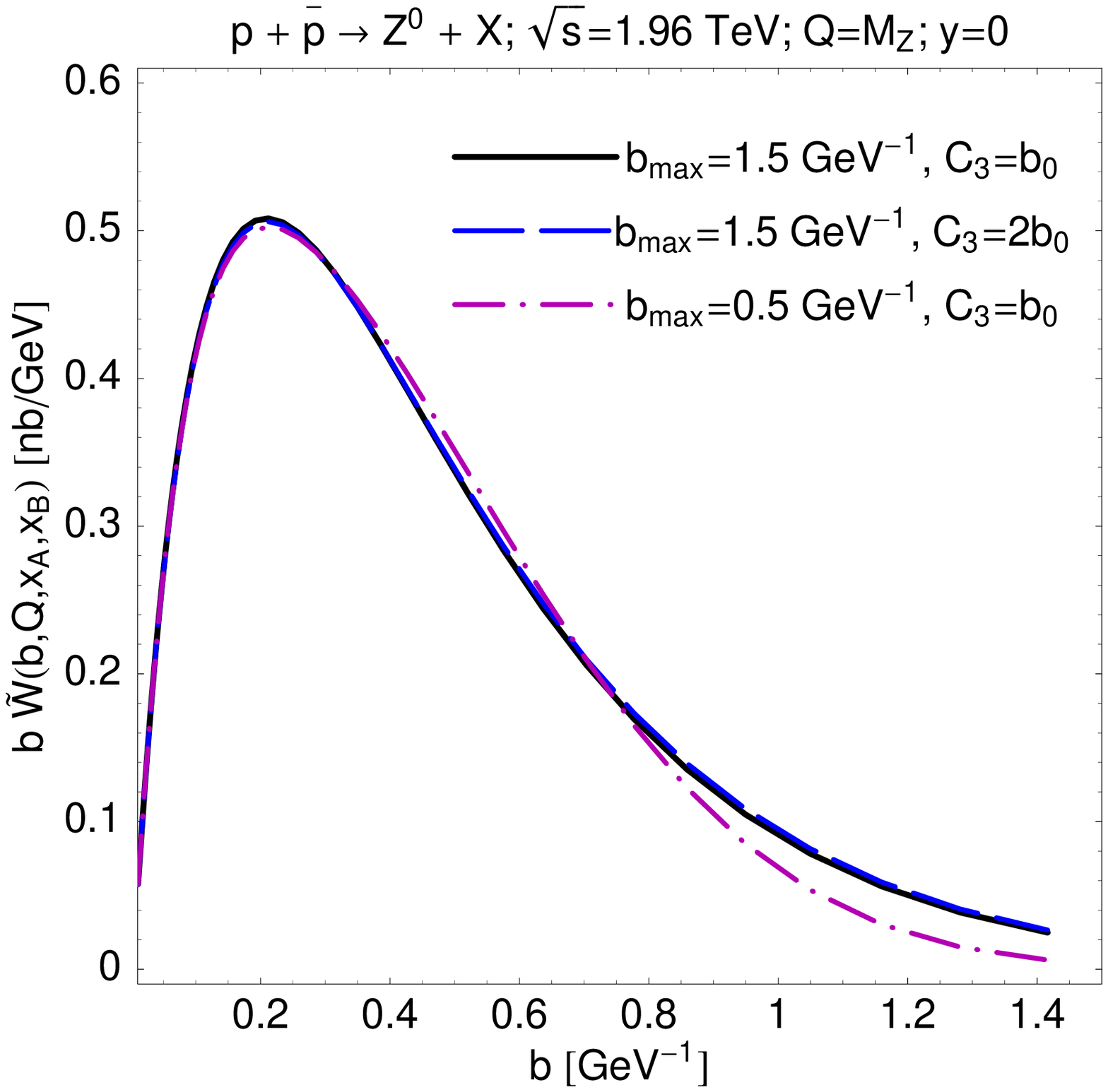}\,\,
\includegraphics[%
  width=0.47\textwidth,
  keepaspectratio]{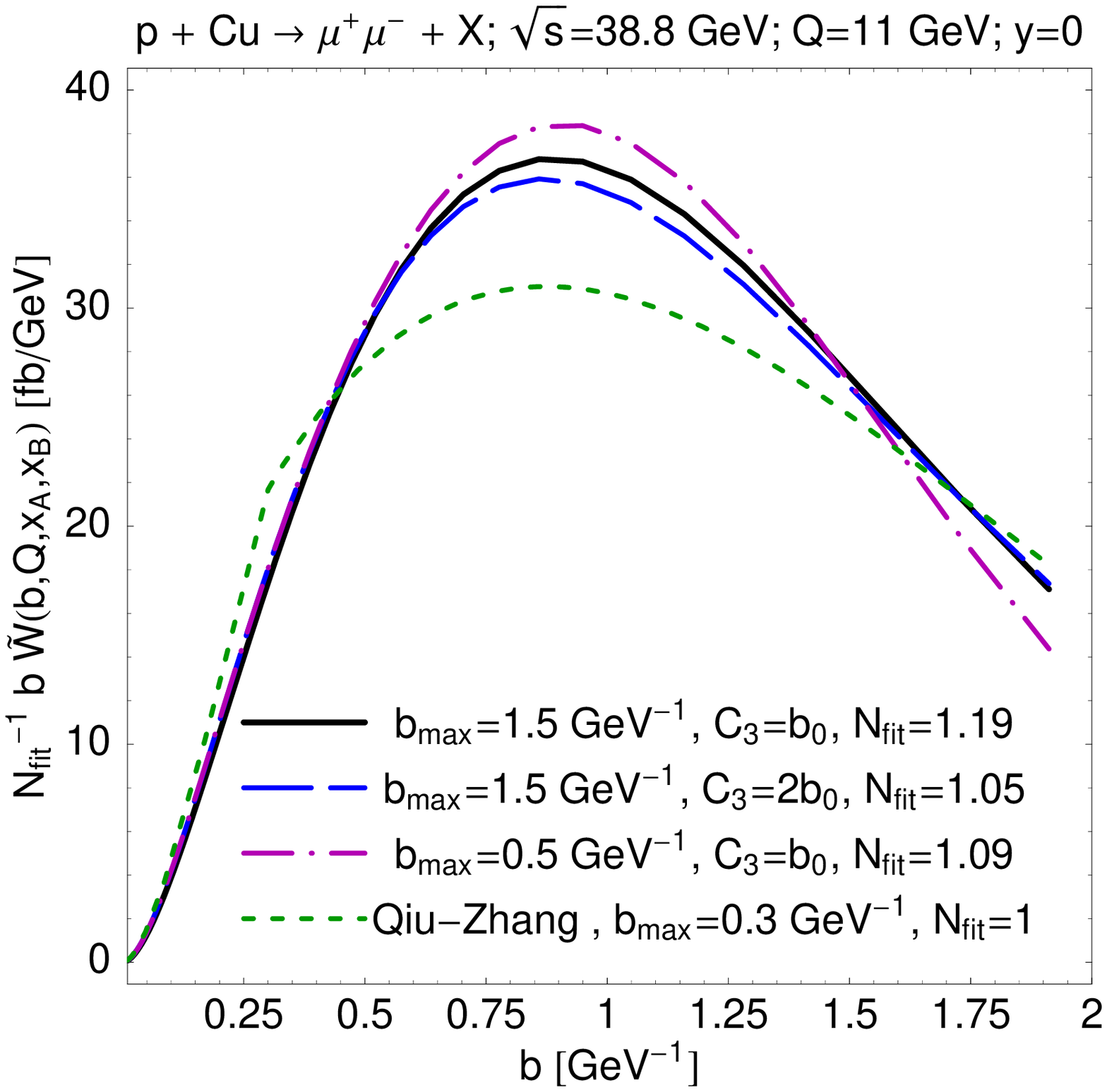}
\end{center}
\caption{The best-fit form factors $b\widetilde W(b)$ in (a) Tevatron Run-2 $Z$ boson production; (b) E605 experiment. In the E605 case, $b\widetilde W(b)$ are divided by the best-fit normalizations $N_{fit}$ for the E605 data, and the form factor in the Qiu-Zhang parametrization \cite{QZ} for $b_{max}^{QZ}=0.3\mbox{ GeV}^{-1}$ is 
also shown.\label{fig:bWb}}
\end{figure}

The magnitudes of $a_1$, $a_{2}$, and $a_{3}$ are reduced
when  $b_{max}$ increases from $0.5$ to $1.5\mbox{ GeV}^{-1}$.
In the whole range $1 \le b_{max} \le 2$ GeV$^{-1}$,
$a_{2}$ agrees with the renormalon analysis estimate.
The coefficient
$a_{3}$, which parametrizes deviations from the linear $\ln Q$ dependence,
is considerably smaller ($<0.05$) than both $a_{1}$ and
$a_{2}$ ($\sim0.2$). 
A reasonable quality of the fit is retained if $a_3$ is
set to zero by hand: $\chi^{2}$ increases by $\approx 5$ in such a fit
above its minimum in the fit with a free $a_3$. In contrast, $\chi^2$ increases
by $>200$ units if $a_3=g_1 g_3$ is set to zero at
$b_{max}=0.5\mbox{ GeV}^{-1}$, as it was noticed in the BLNY study.

The preference for the values of $b_{max}$ between  $1$ and $2 \mbox{
  GeV}^{-1}$ indicates, first, that the data do 
favor the extension of the $b$ range where 
$\widetilde{W}_{LP}(b)$ is approximated by the exact $\widetilde{W}_{pert}(b)$.
In $Z$ boson production, this region extends up 
to $3-4\mbox{ GeV}^{-1}$ as a consequence of the strong
suppression of the large-$b$ tail by the Sudakov exponent.
The fit to the $Z$ data is actually independent of $b_{max}$ within
the experimental uncertainties for $b_{max} > 1\mbox{ GeV}^{-1}$. 
The best-fit form factors $b\widetilde W(b)$ for $b_{max} = 0.5$ and $1.5 
\mbox{ GeV}^{-1}$ in $Z$ boson production are shown in Fig.~\ref{fig:bWb}(a). 
   
In the low-$Q$ Drell-Yan process, continuation of  
$b\widetilde{W}_{pert}(b)$ far
beyond $b \approx 1\mbox{ GeV}^{-1}$ raises objections, since   
$b\widetilde{W}_{pert}(b)$  has a maximum and 
is unstable with respect to
higher-order corrections at $b\approx 1.2-1.5\mbox{ GeV}^{-1}$. 
The dubious large contributions to $\widetilde{W}_{pert}(b)$
in this $b$ region would deteriorate the quality of the fit. The $b_*$
prescription with $b_{max} <\nolinebreak2\mbox{ GeV}^{-1}$ 
reduces the impact of the dubious terms on $\widetilde{W}(b)$: for
  $b_{max}$ small enough, the maximum of $\widetilde{W}_{pert}(b_*)$ 
is only reached at $b \gg 1.2\mbox{ GeV}^{-1}$, 
where it is suppressed by $e^{-{\mathcal F}_{NP}(b,Q)}$. 
The best-fit form factors for the E605 kinematics, divided 
by the best-fit normalizations of the E605 data $N_{fit}$, 
are shown in Fig.~\ref{fig:bWb}(b). 

If a very large $b_{max}$ comparable to $1/\Lambda_{QCD}$ is taken,  
$\widetilde{W}_{LP}(b)$  essentially coincides
with $\widetilde{W}_{pert}(b)$, extrapolated to large $b$ by using the known, 
although not always reliable, dependence of $\widetilde{W}_{pert}(b)$
on $\ln{b}$. Similar, but not identical, 
extrapolations  of $\widetilde{W}_{pert}(b)$ to large $b$
are realized in the models \cite{QZ,KSV}, which describe 
the $Z$ data well, in accord with our own findings. 
In $Z$ boson production, our best-fit $a(M_{Z})=0.85\pm0.10\mbox{ GeV}^2$ 
agrees with $0.8\mbox{ GeV}^2$ found in the extrapolation-based 
models, 
and it is about a third of $2.7\mbox{ GeV}^2$ predicted 
by the BLNY parametrization. 
Our results support the conjecture in \cite{QZ} 
that $a_3$ is small if the exact form of $\widetilde{W}_{pert}(b)$ 
is maximally preserved. To describe the low-$Q$ data, the model~\cite{QZ}
allowed a large discontinuity in the first derivative of  
$\widetilde{W}(b)$ at $b$ equal to the separation
parameter  $b_{max}^{QZ}=0.3-0.5\mbox{ GeV}^{-1}$, where switching
from the exact $\widetilde{W}_{pert}(b)$ to its
extrapolated form occurs [cf. Fig.~\ref{fig:bWb}(b)]. In the revised $b_*$ model, such
discontinuity does not happen, and $\widetilde{W}_{LP}(b)$ is closer to the
exact $\widetilde{W}_{pert}(b)$ in a wider $b$ range at low $Q$ 
than in the model~\cite{QZ}. The two models differ substantially at $b\approx 1\mbox{ GeV}^{-1}$, as seen in Fig.~\ref{fig:bWb}(b).

To summarize, the extrapolation of 
$\widetilde{W}_{pert}(b)$ to $b>\nolinebreak1.5\mbox{ GeV}^{-1}$ is disfavored 
by the low-$Q$ data sets, if a purely Gaussian form of 
${\mathcal F}_{NP}$ is assumed. The Gaussian approximation is  adequate, 
on the other hand, in the $b_*$ model with $b_{max}$ in the range $1-2 \mbox{ GeV}^{-1}$. Here variations in $b_{max}$ are compensated well 
by adjustments in $a_1$, $a_2$, and $a_3$, and the full form factor
$b\widetilde{W}(b)$ stays approximately independent of $b_{max}$.
The best-fit parameters in ${\mathcal F}_{NP}$ are quoted for $b_{max}=1.5\mbox{ GeV}^{-1}$
as $\{ a_{1},a_{2},a_{3}\}=\{0.201\pm0.011,$$0.184\pm0.018,$$-0.026\pm0.007\}\mbox{ GeV}^{2}$
for $C_{3}=b_{0}$, and $\{0.247\pm0.016,$ $0.158\pm0.023,$ $-0.049\pm0.012\}\mbox{ GeV}^{2}$
for $C_{3}=2b_{0}$. In Ref.~\cite{KNlong}, the experimental errors
are propagated into various theory predictions with the help of the Lagrange
multiplier and Hessian matrix methods, discussed, e.g., in Ref.~\cite{CTEQ6}.
We find that the global fit places stricter constraints on ${\mathcal F}_{NP}$
at $Q=M_{Z}$ than the Tevatron Run-1 $Z$ data alone. Theoretical
uncertainties from a variety of sources are harder to quantify, and
they may be substantial in the low-$Q$ Drell-Yan process. In particular,
$\chi^{2}$ for the low-$Q$ data improves by $14$ units when the
scale parameter $C_{3}$ in $\mu_F$ is increased from $b_{0}$ to $2b_{0}$,
reducing the size of the finite-order $\widetilde{W}_{pert}(b)$ at
low $Q$. The best-fit normalizations $N_{fit}$ also vary with $C_3$.
The dependence of the quality of the fit on the arbitrary factorization
scale $\mu_{F}$ indicates importance of ${\mathcal O}(\alpha_{s}^{2})$
corrections at low $Q$, but does not substantially increase uncertainties
at the electroweak scale. Indeed, the ${\mathcal O}(\alpha_{s}^{2})$
corrections and scale dependence are smaller in $W$ and $Z$ production.
In addition, the term $a_{2}\ln Q$, which arises from the soft factor
${\mathcal S}(b,Q)$ and dominates ${\mathcal F}_{NP}$ at $Q=M_{Z}$,
shows little variation with $C_{3}$ {[}cf.~Fig.~\ref{fig:chi2a123}c{]}.
Consequently, the revised $b_*$ model with $b_{max}\approx 1.5\mbox{ GeV}^{-1}$
reinforces our confidence in 
transverse momentum resummation at electroweak scales by
exposing the soft-gluon origin and universality of the dominant
nonperturbative terms at collider energies.\\
{\small}\\
We thank C.-P. Yuan for his crucial contribution to the setup of the
fitting program, and T. Londergan, A. Szczepaniak, S. Vigdor, and CTEQ 
members for the helpful discussions. This work was supported 
by the NSF grants PHY-0100348 and PHY-0457219, and DOE grant W-31-109-ENG-38.

\end{document}